\documentclass[10pt]{article}
\usepackage[font={small}, labelsep=period, indention=0.5in]{caption}
\usepackage[OE]{express}
\usepackage{subfig}
\usepackage{float}

\usepackage{graphicx}
\usepackage{lipsum} 
\usepackage{booktabs}
\usepackage{tabularx}
\usepackage{array}
\newcolumntype{L}[1]{>{\raggedright\let\newline\\\arraybackslash\hspace{0pt}}m{#1}}
\newcolumntype{C}[1]{>{\centering\let\newline\\\arraybackslash\hspace{0pt}}m{#1}}
\newcolumntype{R}[1]{>{\raggedleft\let\newline\\\arraybackslash\hspace{0pt}}m{#1}}

\begin{document}

\title{Arrayed waveguide grating spectrometers for astronomical applications: New results}

\author{Pradip Gatkine,\authormark{1*} Sylvain Veilleux,\authormark{1,2} Yiwen Hu,\authormark{3} Joss Bland-Hawthorn,\authormark{4} and Mario Dagenais\authormark{3}}

\address{\authormark{1}Department of Astronomy, University of Maryland, College Park, Maryland 20742, USA\\
\authormark{2}Joint Space-Science Institute, University of Maryland, College Park, Maryland 20742, USA \\
\authormark{3}Department of Electrical and Computer Engineering, University of Maryland, College Park, Maryland 20742, USA\\\authormark{4}Sydney Institute for Astronomy, School of Physics, University of Sydney, NSW 2006, Australia}

\email{\authormark{*}pgatkine@astro.umd.edu} 
\homepage{https://www.astro.umd.edu/$\sim$pgatkine/} 


\begin{abstract}
One promising application of photonics to astronomical instrumentation is the miniaturization of near-infrared (NIR) spectrometers for large ground- and space-based astronomical telescopes. Here we present new results from our effort to fabricate arrayed waveguide grating (AWG) spectrometers for astronomical applications entirely in-house. Our latest devices have a peak overall throughput of $\sim$23\%, a spectral resolving power ($\lambda/\delta\lambda$) of $\sim$1300, and cover the entire H band (1450$-$1650 nm) for Transverse Electric (TE) polarization. These AWGs use a silica-on-silicon platform with a very thin layer of Si$_3$N$_4$ as the core of the waveguides. They have a free spectral range of $\sim$10 nm at a wavelength of $\sim$1600 nm and a contrast ratio or crosstalk of about 2\% ($-$17 dB). Various practical aspects of implementing AWGs as astronomical spectrographs are discussed, including the coupling of the light between the fibers and AWGs, high-temperature annealing to improve the throughput of the devices at $\sim$1500 nm, cleaving at the output focal plane of the AWG to provide continuous wavelength coverage, and a novel algorithm to make the devices polarization insensitive over a broad band. These milestones will guide the development of the next generation of AWGs with wider free spectral range and higher resolving power and throughput. 
\end{abstract}



\section{Introduction}

The study of the first billion years of the universe, corresponding to cosmological redshifts $z \sim 6-12$ due to the expansion of the universe, is crucial to understand phenomena such as galaxy formation, the ionization of the intergalactic medium, and the formation and evolution of supermassive black holes. The rest-frame ultra-violet light coming from sources in this range of cosmological redshifts is shifted to the J and H bands ($1.15 - 1.4$ $\mu m$ and $1.45 - 1.7$ $\mu m$, respectively) in the near-infrared (NIR). Therefore, it is of astrophysical interest to study the NIR spectra of these distant and faint sources, for which large telescopes such as the Keck 10-meter telescopes are required. The next generation of ground-based extremely large telescopes (ELTs) in the optical and NIR will have diameters in the range of 20$-$40 meters. This necessitates the development of suitable seeing-limited spectroscopic instrumentation for astrophysical studies \cite{bland2006instruments}. The size of the optical components in a conventional spectrograph scales roughly with the telescope diameter $D$, hence the volume, mass, and cost of the instrument scale roughly as $D^{3}$ \cite{bland2006instruments}. This highlights the need for innovation in building instruments for the upcoming ELTs. 

The application of photonic technologies \cite{pervez2010photonic,roelkens2013silicon,pathak2014comparison} to astronomical spectroscopy is a promising approach to miniaturize the next-generation spectrometers for large telescopes\cite{harris2013applications}. This is attained mainly by leveraging the two-dimensional photonic structures on a chip\cite{chaganti2006simple, bland2009astrophotonics,allington2010astrophotonic,subramanian2015silicon}, thus reducing the size of spectroscopic instrumentation to a few centimeters and the weight to a few hundreds of grams. Such integrated photonic spectrometers are also more amenable to complex light manipulation and massive multiplexing, cheaper to mass produce, easier to control, and much less susceptive to vibrations and flexures than conventional astronomical spectrographs with similar specifications (resolution, efficiency, and operating wavelength range) \cite{cvetojevic2010miniature}. In this paper, we explore one such photonic technology, the arrayed waveguide gratings (AWGs), designed to be implemented as an astronomical spectrometer in the NIR H band.

\section{Arrayed waveguide gratings}

In many ways, arrayed waveguide gratings are analogous to conventional grating spectrographs (see Fig. 1 in \cite{gatkine2016development}). In a conventional spectrograph, the light source illuminates the grating through an input lens, the grating creates a path difference between different light paths and the output lens focuses the emergent light on the focal plane. In an AWG, these actions take place on a chip, where the single-mode waveguides guiding the light serve as different light paths.  The light from the source is carried by a single-mode waveguide and launched into an input lens, called input free propagation region (FPR), where it illuminates an array of waveguides (similar to illuminating a grating). These waveguides are constructed to introduce a constant path difference between the adjacent waveguides, according to the spectral order. The light from the array of waveguides is focused in the output FPR, with different wavelengths interfering constructively at different spatial locations along the focal plane. The output waveguides carry this dispersed light for measurement. The various components of our AWG are shown in Fig. ~\ref{fig:AWG_CAD}.         

A detailed theory of AWG design is described in the pioneering work on AWG devices \cite{smit1996phasar}. Traditionally, the AWG devices are used for wavelength division multiplexing (WDM) in telecommunication industry around a wavelength of 1550$\pm$50 nm. But in principle, the same theory can also be used for spectroscopic purposes. In particular, some of the recent work towards making low-loss AWG devices \cite{bauters2010ultra, dai2011low, akca2011high} demonstrates the usability of these techniques for NIR spectroscopy. There have also been successful preliminary tests of using modified commercial AWGs for astronomical spectroscopy \cite{cvetojevic2012developing, cvetojevic2012first}. AWGs, along with other advances in the field of astrophotonics, such as photonic lanterns \cite{leon2010photonic,thomson2011ultrafast, birks2015photonic} to convert multimode fibers to single mode fibers,  Bragg gratings (in fibers \cite{othonos1997fiber,meltz1989formation,trinh2013gnosis, lindley2014demonstration} as well as on chips \cite{zhu2016arbitrary}) to suppress the unwanted atmospheric OH-emission background (in the NIR), and high-efficiency fiber bundles for directly carrying the light from the telescope focal plane \cite{lawrence2012hector}, offer a complete high-efficiency miniaturized solution for astronomical spectroscopy in the NIR. This solution has potential applications for future ground-, balloon- and space-based telescopes.

The technical requirements for our AWG spectrograph are driven by the science goals. Our main science goal here is the study of faint sources at high cosmological redshifts ($z \gtrsim 6$) to probe the first billion years of the universe. This requires a spectral resolving power ($\lambda$/$\delta\lambda$) of at least $\sim$1500 in the H band to measure the redshifts of these sources and distinguish between different absorption lines produced by the intervening material between the observers and these sources \cite{vreeswijk2006low,salvaterra2015high}. Also, a wide spectral range (preferably both J and H bands) is necessary to ensure that these absorption lines fall within the band-pass at cosmological redshifts larger than 6. The throughput of the spectrograph should at least be comparable to that of the conventional astronomical spectrographs (from slit to detector, typically $\sim$20\%, although this depends on the specific instrument and configuration \cite{GNIRS,MOIRCS}).   

In our previous paper \cite{gatkine2016development}, we demonstrated AWG spectrometers in the H band with a resolving power of 1250 and a peak overall throughput of 13\% for transverse electric (TE) polarization. For practical implementation as a competitive astronomical spectrometer, this throughput needs to be improved. Also, the overlapping spectral orders of AWG (at the output focal plane) need to be cross-dispersed to extract the final spectrum, which requires the focal plane of the AWG to be exposed to the cross-dispersion optics \cite{cvetojevic2012developing,cvetojevic2012first}. This necessitates cleaving the AWG at its focal plane. The present paper addresses all of these issues. We first describe the procedures used to design, fabricate, and characterize the new AWGs. Two H-band AWG devices are fabricated to demonstrate the relevant techniques of coupling-taper optimization, annealing, and cleaving at the focal plane. A new way to design a polarization-insensitive AWG is discussed next. Future avenues of research to further improve the throughput of our devices and allow us to expand the wavelength range to the J band are discussed in the last section.

\section{Methods}

Two new AWGs are presented in this paper. Their characteristics are summarized in Table \ref{tab:awg_summary}. The main difference between these two devices is the use of output waveguides in AWG \#1, while AWG \#2 has a cleaved, open-faced output. In this section, we discuss in detail the design, fabrication, and characterization methods of AWG \#1. Most of this discussion also applies to AWG \#2. The differences are discussed in detail in the next section. 

\begin{table}[ht]
\caption{Summary of the characteristics of the two AWGs.} 
\label{tab:awg_summary}
\begin{center}       
\begin{tabular}{|C{4.5cm}|C{3.75cm}|C{3.75cm}|} 
\hline
\rule[-1ex]{0pt}{3.5ex}   & AWG \#1 & AWG \#2 \\
\hline
\rule[-1ex]{0pt}{3.5ex} Waveguide cross-section & 2.0 $\times$ 0.1 $\mu m$ & 2.0 $\times$ 0.1 $\mu m$ \\
\hline
\rule[-1ex]{0pt}{3.5ex}  Number of waveguides & 34 & 34 \\
\hline
\rule[-1ex]{0pt}{3.5ex}  FPR length & 200 $\mu m$ & 200 $\mu m$ \\
\hline
\rule[-1ex]{0pt}{3.5ex}  $\Delta L$ & 172 $\mu m$ & 172 $\mu m$ \\
\hline
\rule[-1ex]{0pt}{3.5ex}  Separation between waveguides at array-FPR interface & 6 $\mu m$ & 6 $\mu m$ \\
\hline
\rule[-1ex]{0pt}{3.5ex}  Output waveguide spacing & 6 $\mu m$ & Cleaved, open faced \\
\hline
\rule[-1ex]{0pt}{3.5ex} Inputs & Fiber-coupling tapers & Fiber-coupling tapers \\
\hline
\rule[-1ex]{0pt}{4.5ex} Outputs & Fiber-coupling tapers & Cleaved, open faced \\
\hline
\rule[-1ex]{0pt}{4.5ex} Footprint & 16 mm $\times$ 7 mm & 12 mm $\times$ 8 mm \\

\hline
\end{tabular}
\end{center}
\end{table}

\subsection{Design} \label{design}

\begin{figure}[htbp]
\subfloat[]{\includegraphics[height=3.4cm]{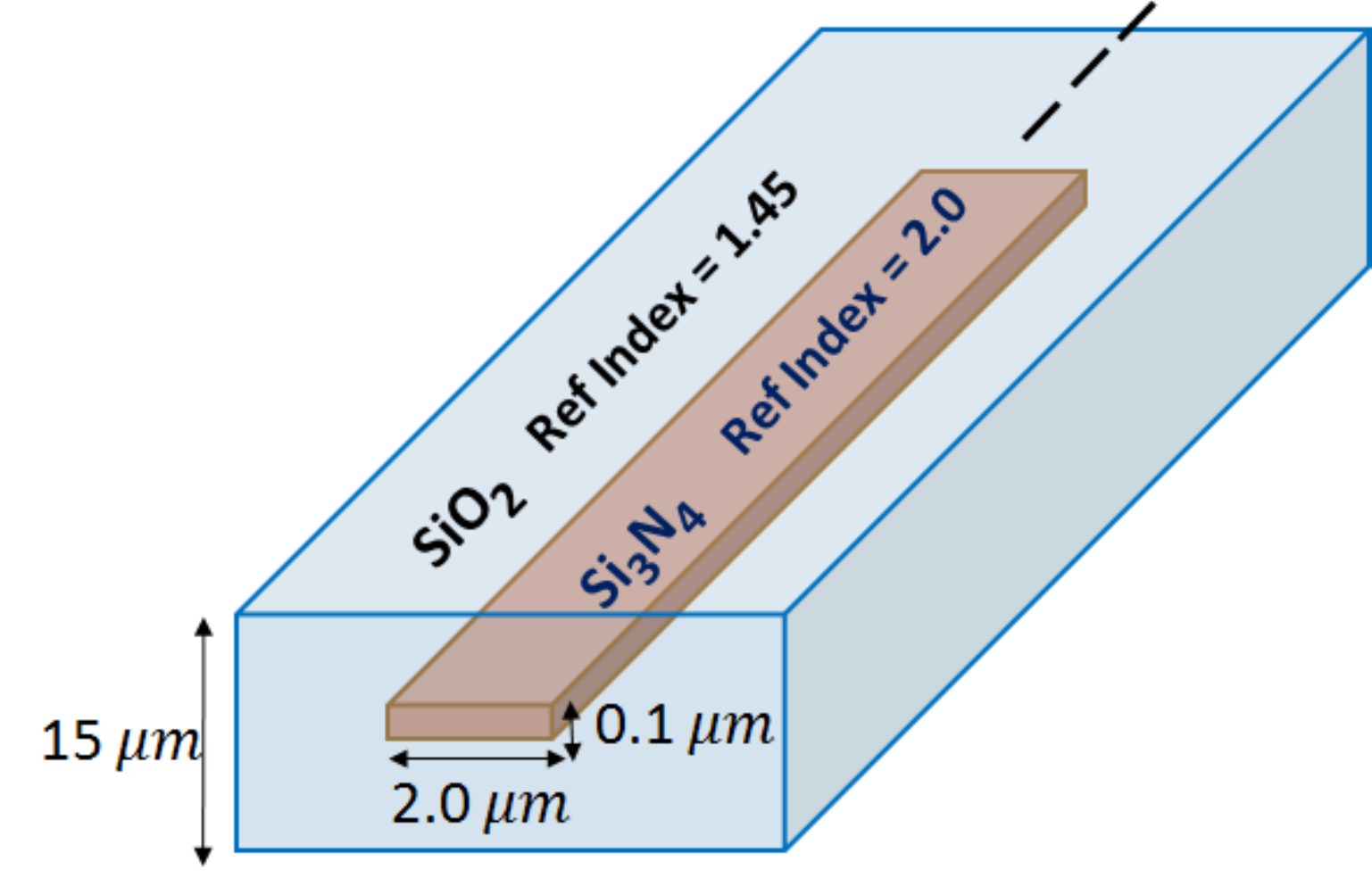}}
\subfloat[]{\includegraphics[height=3.4cm]{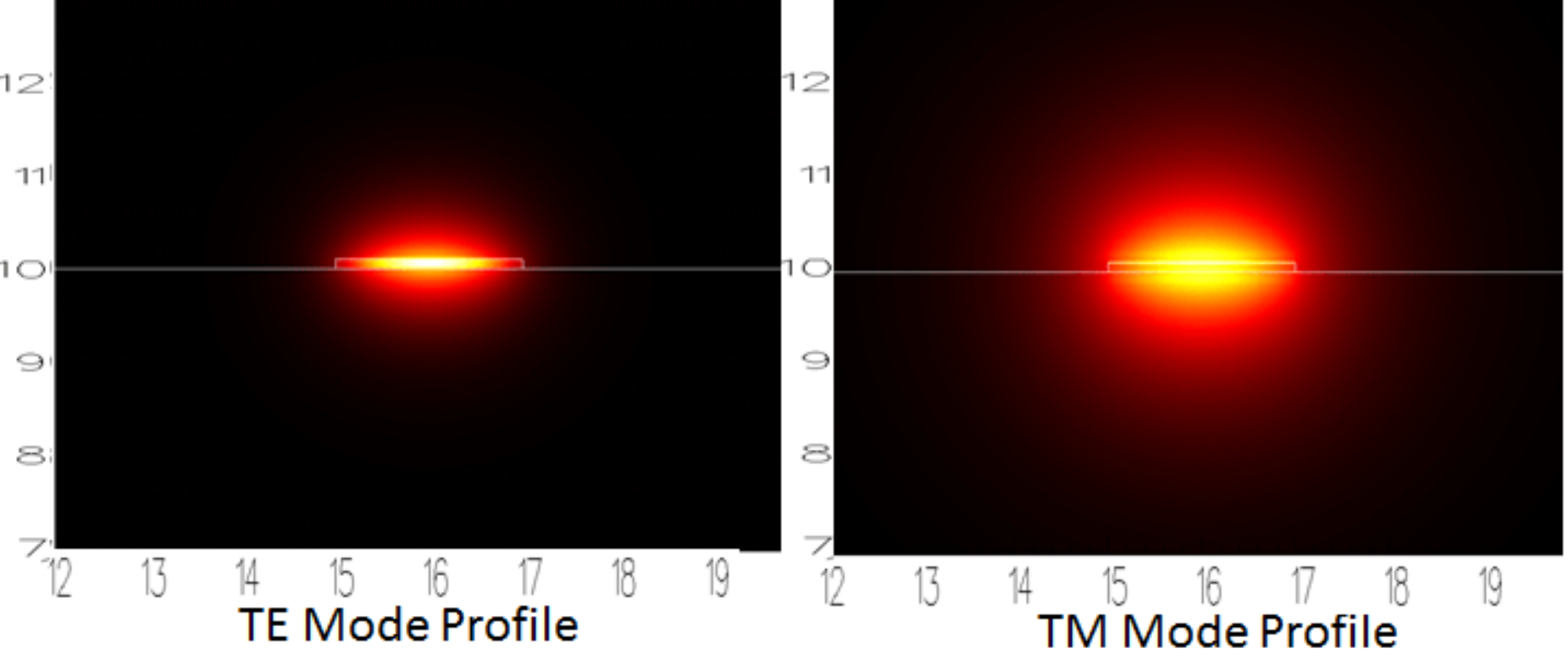}}
\caption{a) The Si$_3$N$_4$/SiO$_2$ waveguides used in AWG \#1 and \#2. b) Mode profile for TE ($n_{eff}$ = 1.4659) and TM polarizations ($n_{eff}$ = 1.4473). Note that the TM mode is weakly confined. }
\label{fig:Waveguide_geometry} 

\end{figure}

The selection of the waveguide material is crucial for building low-loss AWGs. In the recent past, Si$_3$N$_4$ has been proven to be one of the best suited materials for low-loss photonic devices \cite{dai2012passive}. Therefore, we use Si$_3$N$_4$ (refractive index $\sim$ 2.0) waveguides buried in SiO$_2$ (refractive index $\sim$ 1.45) as shown in Fig.~\ref{fig:Waveguide_geometry} for low on-chip transmission losses\cite{bauters2010ultra, gatkine2016development}. The most important sources of losses in the AWGs are: a) the coupling loss (fiber to chip and vice versa), b) the sidewall scattering loss due to sidewall roughness and micro-cracks, c) the bending loss due to radiative loss (especially in weakly confined waveguide modes), and d) the absorption loss due to absorption features of the material and/or inadvertent impurities. In our previous work \cite{gatkine2016development}, we focused on sidewall scattering and bending losses by using a $2.8 \times 0.1 \mu$m waveguide geometry and demonstrated a high on-chip throughput (peak $\sim$80\%), but a relatively modest overall throughput ($\sim$13\%). The present paper addresses this problem of low overall throughput.

The adopted geometry of the waveguides is shown in Fig.\ 1. A thickness of 0.1 $\mu m$ is selected for several reasons. Reproducibility of the fabrication process is an important issue. It is easier to control the actual deposited thickness of the layer of nitride (within a tolerance of 5-10\%) if it is $\sim$0.1 $\mu m$ or thicker. However, the sidewall scattering loss is proportional to the sidewall area and hence the height of the waveguide. So a 0.1 $\mu m$ thickness provides a balance between deposition non-uniformity and sidewall-roughness induced in the etching process\cite{bauters2010ultra,bauters2011planar,bauters2011ultra}. Moreover, with a thickness of 0.1 $\mu m$, we can use a relatively narrow waveguide and achieve a mode-confinement factor (13.3\% for TE polarization) that is similar to that of a wider and thinner waveguide (eg. \cite{bauters2010ultra}). This reduces the chip size and makes it easier to fabricate the devices with precise electron-beam lithography (e-beam writing time scales with the writing area). We select the waveguide width to be 2 $\mu m$, different from our previous $2.8$ $\mu m$ design, because it helps in packing the same number of arrayed waveguides in a smaller area without degrading the confinement. With a relatively smaller footprint, it is easier to use a larger radius of curvature (R$_{min}$ = 2.5 mm) for the curved waveguides, thus preventing the curvature loss \cite{bauters2010ultra}. In contrast to a waveguide with a square cross-section (such as 0.4 $\times$ 0.4 $\mu m$), this high-aspect ratio waveguide is easier to fabricate; it has a greater tolerance for width errors and provides better etch-depth uniformity due to the thin structure. This waveguide geometry is also better matched to the taper geometry used to improve the fiber-AWG coupling efficiency \cite{zhu2016ultrabroadband}, as described in section 4.1.  

We calculated the mode profile for the 2 $\mu m$ $\times$ 0.1 $\mu m$ waveguide using a full-vectorial finite difference method simulation in FIMMWAVE software \cite{FIMMWAVE} and confirmed the single-mode nature of the waveguide over a wide range of wavelengths ($\lambda$ > 1000 nm). The simulated mode profiles for Transverse Electric (TE) and Transverse Magnetic (TM) polarizations at a wavelength of 1550 nm and the geometry of the waveguides are shown in Fig.~\ref{fig:Waveguide_geometry}. The index contrast of the waveguide is 23.7\%. A moderate spectral order (m) of 165 (at $\lambda$ = 1600 nm) is used in this design to maximize the free spectral range and obtain the desired resolution, while keeping the required number of waveguides small (34 waveguides), thus reducing the electron-beam lithography time.

\begin{figure} [ht!]
   \begin{center}
   \begin{tabular}{c} 
   \includegraphics[height=7.5cm]{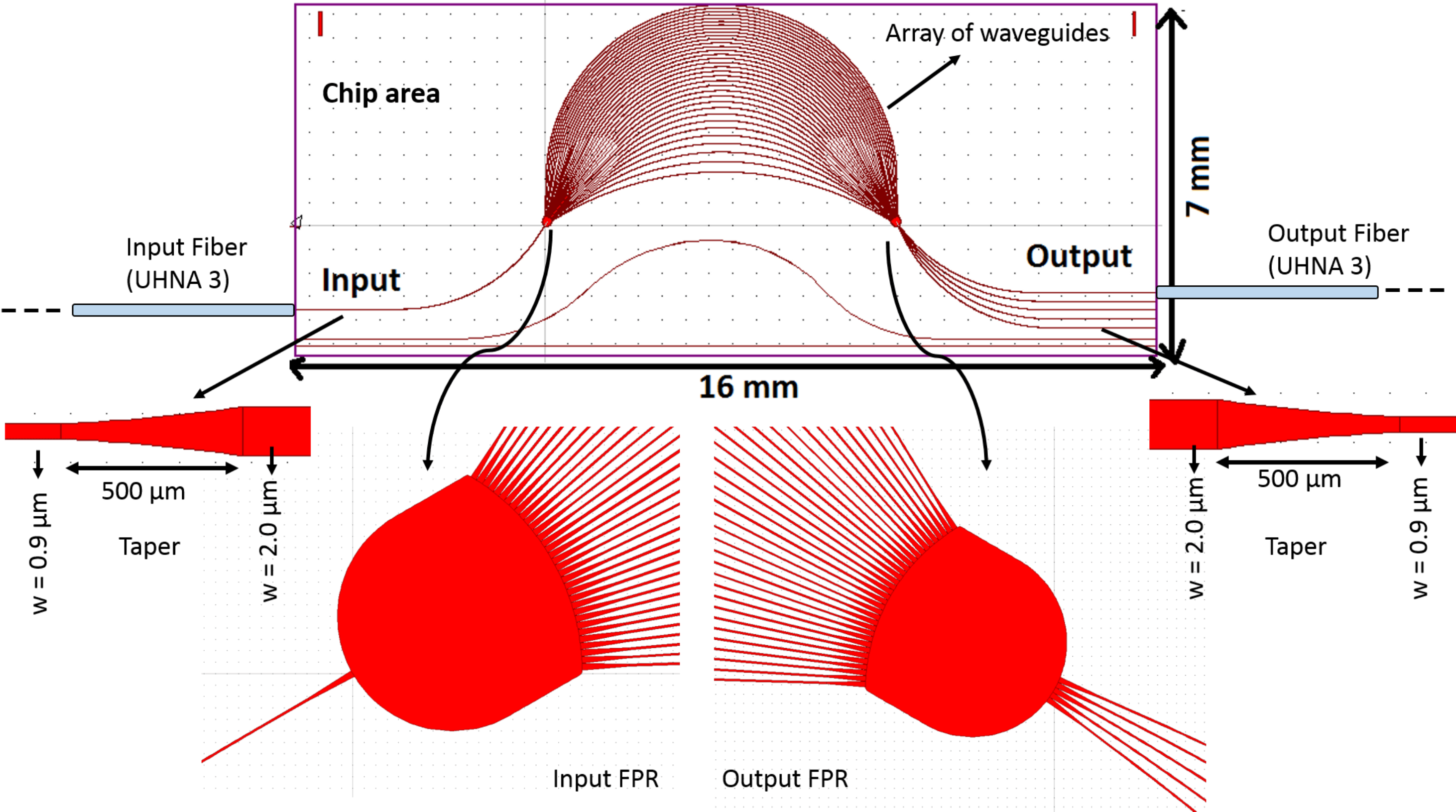}
   \end{tabular}
   \end{center}
   \caption{CAD of AWG \#1. Note the vertical cleaving marks near the top corners of the chip to aid cleaving the edges to expose the optical quality cross-section of the waveguides for fiber coupling. The extra waveguides at the bottom are reference waveguides for calibration. The AWG has a small footprint of only 16mm $\times$ 7mm. The actual writing area is 11.5 mm$^{2}$, thus making it suitable for e-beam lithography. The AWG input, output and the reference waveguides have on-chip coupling tapers as a continuation of the waveguides, shown in left and right insets. UHNA3 fibers are used for characterization by butt-coupling one by one to the tapers. A zoomed-in version of the input and output FPRs are shown at the bottom for clarity.}
   { \label{fig:AWG_CAD} 
}
   \end{figure}

With these initial parameters, we designed an AWG for the H band (1450$-$1650 nm) using the design algorithm prescribed in \cite{smit1996phasar}. To design the AWG, we used a central wavelength of 1550 nm and a desired spectral channel spacing of 1.6 nm. The H band is covered in 23 spectral orders, with a free spectral range (FSR) varying from 8 nm at a wavelength of 1450 nm to 10 nm at 1650 nm ($FSR$ $\approx$ $\lambda_{0}/m$ $\times$ $n_{eff}/n_{group}$, where $n_{eff}$ and $n_{group}$ are the effective and group indices of refraction, respectively). Five output waveguides are used to sufficiently sample the output FPR for AWG characterization. The AWG has a total of 34 waveguides in the array to ensure adequate sampling of the input free propagation region (FPR). The length difference ($\Delta L$) between adjacent waveguides of the array is 172 $\mu m$ ($\Delta L$ $=$ $m$ $\times$ $\lambda_{0}/n_{eff}$). The spatial channel spacing at the output FPR is 6 $\mu m$. The length of the FPR is 200 $\mu m$. The tapers at the input and output FPRs are linear, with a length of about 30 $\mu m$ and a width equal to the waveguide separation at the FPR (hence, the taper width at the FPR is 6 $\mu m$). This ensures optimal transmission of light from the FPR to the array of waveguides and vice versa. Such a geometry of touching tapers requires a precise fabrication which is made possible with electron beam lithography (within a tolerance of 10 nm). The minimum bending radius in the layout is 2.5 mm to reduce the curvature loss. Some straight and curved reference waveguides are also fabricated below the AWG to characterize the on-chip loss of the AWG. A coupling taper was added (on the chip) for all of the devices presented in this paper to optimize the coupling between the fiber (UHNA3) and the waveguide (further details are in section 4.1). The AWG layout is shown in Fig. \ref{fig:AWG_CAD}  

\subsection{Fabrication} \label{fab}

\begin{figure} [ht]
   \begin{center}
   \begin{tabular}{c} 
   \includegraphics[height=3.0cm]{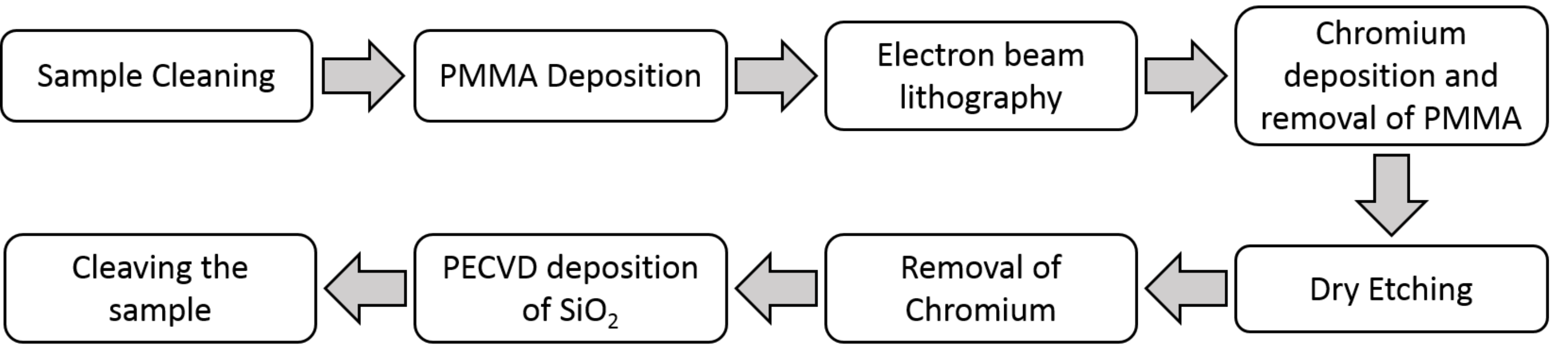}
   \end{tabular}
   \end{center}
   \caption[Fabrication_process] 
   { \label{fig:Fabrication_process} 
 Fabrication sequence of AWGs \cite{gatkine2016development}.}
   \end{figure}

The fabrication sequence was the same as the one used in our earlier paper \cite{gatkine2016development}. For completeness, it is summarized in Fig.~\ref{fig:Fabrication_process} and briefly described here again. A silicon wafer, pre-deposited with 10 $\mu m$ layer of thermal silica (SiO$_2$), was used for the fabrication of the AWG chip. A 0.1 $\mu m$ thick layer of silicon nitride (Si$_3$N$_4$) was deposited on top of that using LPCVD (Low Pressure Plasma Enhanced Deposition). A smaller sample of the wafer (30 $\times$ 15 mm) was obtained by dicing the wafer. The process sequence followed to fabricate the buried silicon nitride AWG device was: spin-coating the photoresist on the sample, electron-beam lithography (moving e-beam to write the pattern), electron-beam chromium metal deposition, chromium lift-off (leaving only the chromium mask for etching), reactive ion etching (RIE) to a depth of 0.1 $\mu m$, chromium etching to dissolve the mask and finally, PECVD (Plasma Enhanced Chemical Vapor Deposition) of 6 $\mu m$ of SiO$_2$ as the upper cladding layer of the device. After fabrication, the sample was cleaved at precise locations from left and right (along the crystal plane of the chip) to expose facets of the input and output waveguides for coupling the light. The facets were then examined for optical quality. 

\subsection{Characterization} \label{char}

\begin{figure} [htp]
   \begin{center}
   \begin{tabular}{c} 
\includegraphics[clip,height=5cm]{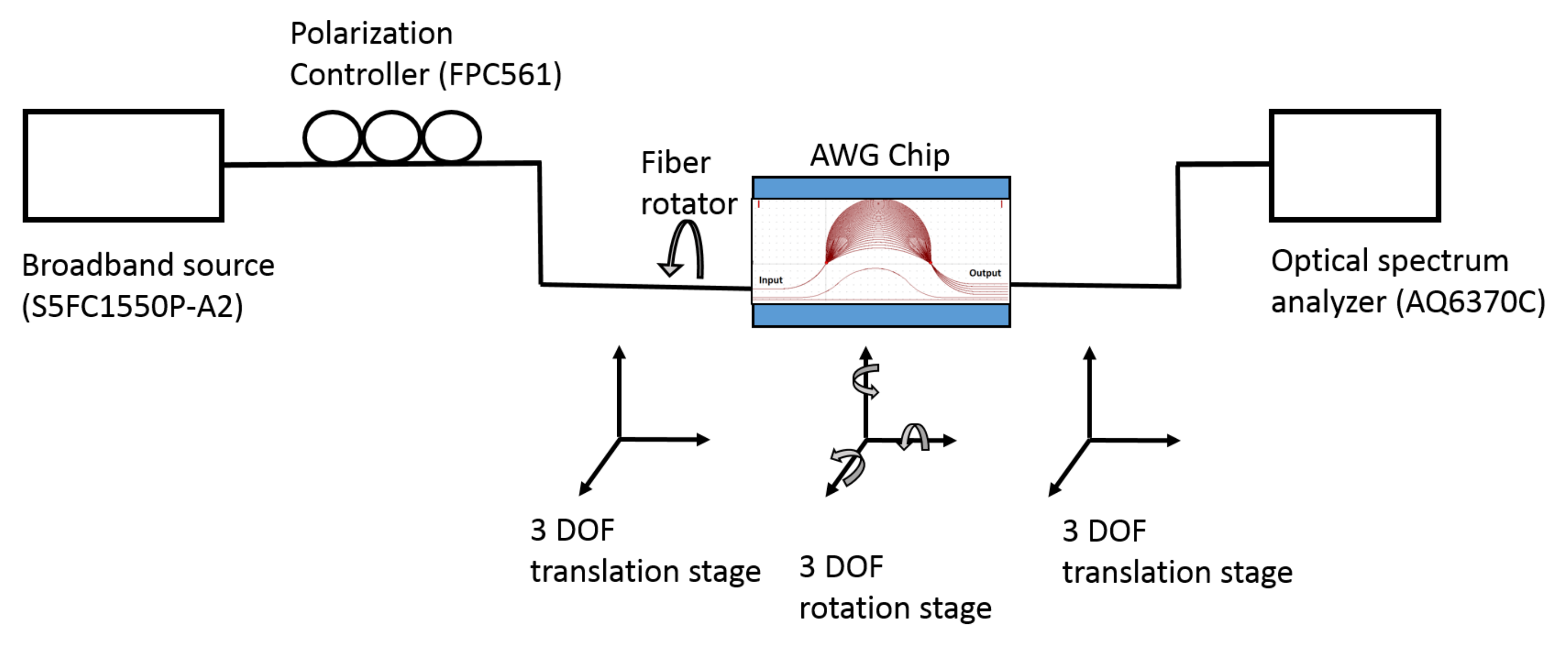}
\\
\includegraphics[clip,height=5cm]{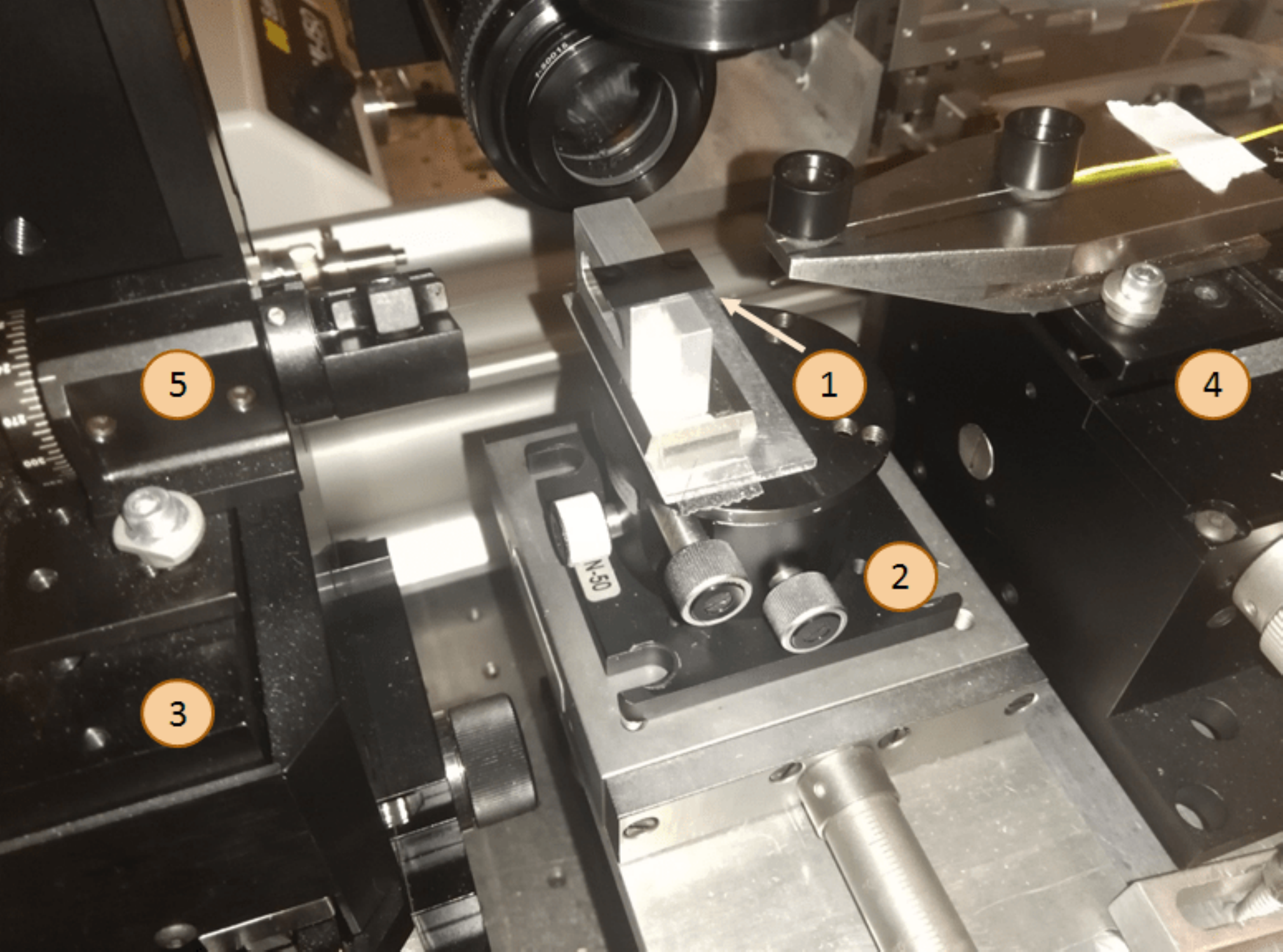}
   \end{tabular}
   \end{center}
   \caption[Characterization setup] 
   { \label{fig:Characterization_stage} Top panel: A schematic of the setup used for AWG characterization. Bottom panel: The AWG sample (label 1) is mounted in the center on top of a tip-tilt-rotation mount (label 2). The input and output fibers are mounted on 3-axis stages with 10 nm precision (left:3 and right:4). The input fiber is mounted on a fiber rotator (label 5), which, along with the polarization controller, allows for polarization tuning.
}
   \end{figure}

To characterize the transmission response of the AWG, we used a polarization maintaining broadband superluminescent diode source by Thorlabs (S5FC1550P- A2, amplified spontaneous emission power of 2.5 mW) operating in a waveband of 1450 nm $-$ 1650 nm (corresponding to 20 dB width). An Optical Spectrum Analyzer (OSA, YOKOGAWA AQ6370C) with a dynamic range of 72 dB across the H band was used to analyze the signal. Ultra-high numerical aperture (UHNA3) fibers with a typical mode-field diameter ($1/e^{2}$ diameter) of 4.1 $\mu m$ and a numerical aperture of 0.35 were used to carry the signal from the broadband source to the AWG and out to the OSA. These fibers are single mode over the entire range of H band. In the characterization setup, a UHNA3 fiber was connected to the broadband source and butt-coupled to the AWG input waveguide through a fiber polarization controller (Thorlabs, FPC561) and a fiber rotator. The combination of polarization controller and fiber rotator were used to control the polarization since UHNA3 is not a polarization maintaining fiber. Another UHNA3 fiber was butt-coupled to one of the output waveguides and connected to the OSA. The optical butt-coupling of the fibers to the facet of the chip was done by carefully aligning the fibers and the AWG chip using a 9 degree-of-freedom alignment setup. An index matching solution (index = 1.45) was used to minimize any reflections at the fiber-waveguide interface.  The characterization setup is shown in Fig.~\ref{fig:Characterization_stage}.

The broadband light source was measured to be steady as a function of time within 0.05 dB. The output fiber was coupled to each of the AWG output channels one-by-one and the 
transmission response (end-to-end) of each channel was recorded with the OSA. Similarly, the transmission response of the curved reference waveguide was obtained by butt-coupling the fibers to it. This transmission response (fiber-AWG-fiber) was normalized to the input power spectrum (fiber-fiber) to obtain the `overall AWG response' (including the coupling efficiency). The overall AWG response was further normalized to that of the curved reference waveguide to isolate the `on-chip response' of the AWG.

\section{Results}\label{Key Results}

In this section, we describe our results emphasizing three critical aspects of the new AWG devices: 1. fiber-coupling tapers, 2. annealing, and 3. cleaving at the output FPR. All of the results are measured for TE polarization.

\subsection{AWG \#1: fiber-coupling taper}

\begin{figure} [ht]
   \begin{center}
   \begin{tabular}{c} 
   \includegraphics[height=4cm]{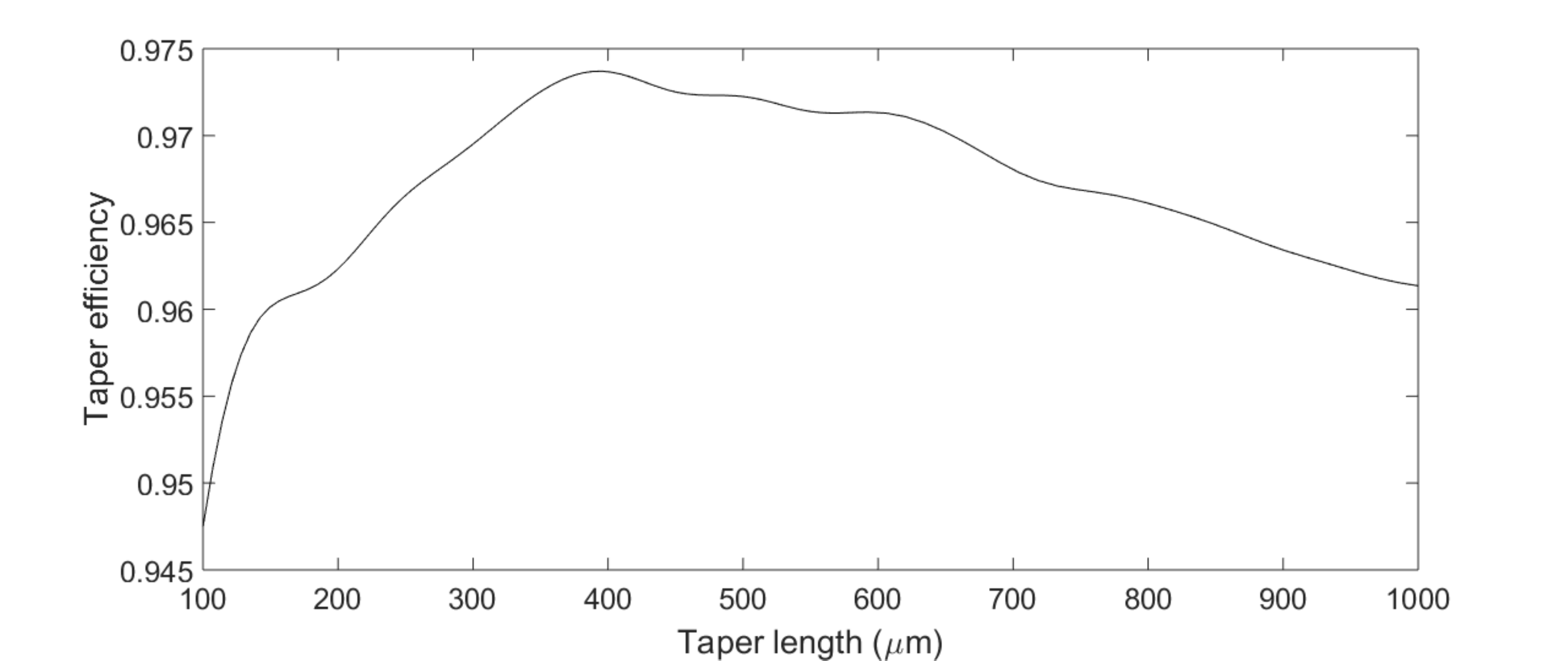}
   \end{tabular}
   \end{center}
   \caption[Taper efficiency] 
   { \label{fig:Taper simulation} Simulation of taper conversion efficiency as a function of length for a linear taper. Here, we have also taken into account an estimated propagation loss for the taper of 1.5 dB/cm.
}
   \end{figure}

Zhu et al. \cite{zhu2016ultrabroadband} demonstrated a coupling efficiency of >90\% (1450 - 1650 nm) using UHNA3 fiber and 0.9 $\mu m$ $\times$ 0.1 $\mu m$ waveguide geometry for TE polarization. In this paper, we present a taper geometry to convert a weakly guided fiber-side waveguide cross-section (0.9 $\mu m$ $\times$ 0.1 $\mu m$) to a relatively strongly guided AWG-side waveguide cross-section (2.0 $\mu m$ $\times$ 0.1 $\mu m$) to improve the coupling efficiency and thus, the overall transmission of the AWG. In Fig. \ref{fig:Taper simulation} we show the simulated conversion efficiency of a linear taper for this configuration as a function of the taper length for TE mode. Here, we have also added an estimated propagation loss for the taper ($\sim$1.5 dB/cm \cite{zhu2016ultrabroadband}) to find the optimal taper length. Thus, we selected a length of 500 $\mu m$ for the taper. This taper geometry is shown in Fig. \ref{fig:AWG_CAD} and is used on both the input and output sides of the AWG and reference waveguides.

\begin{figure} [htb!]
   \begin{center}
   \begin{tabular}{c} 
 	\centering
	\includegraphics[height=15.5cm]{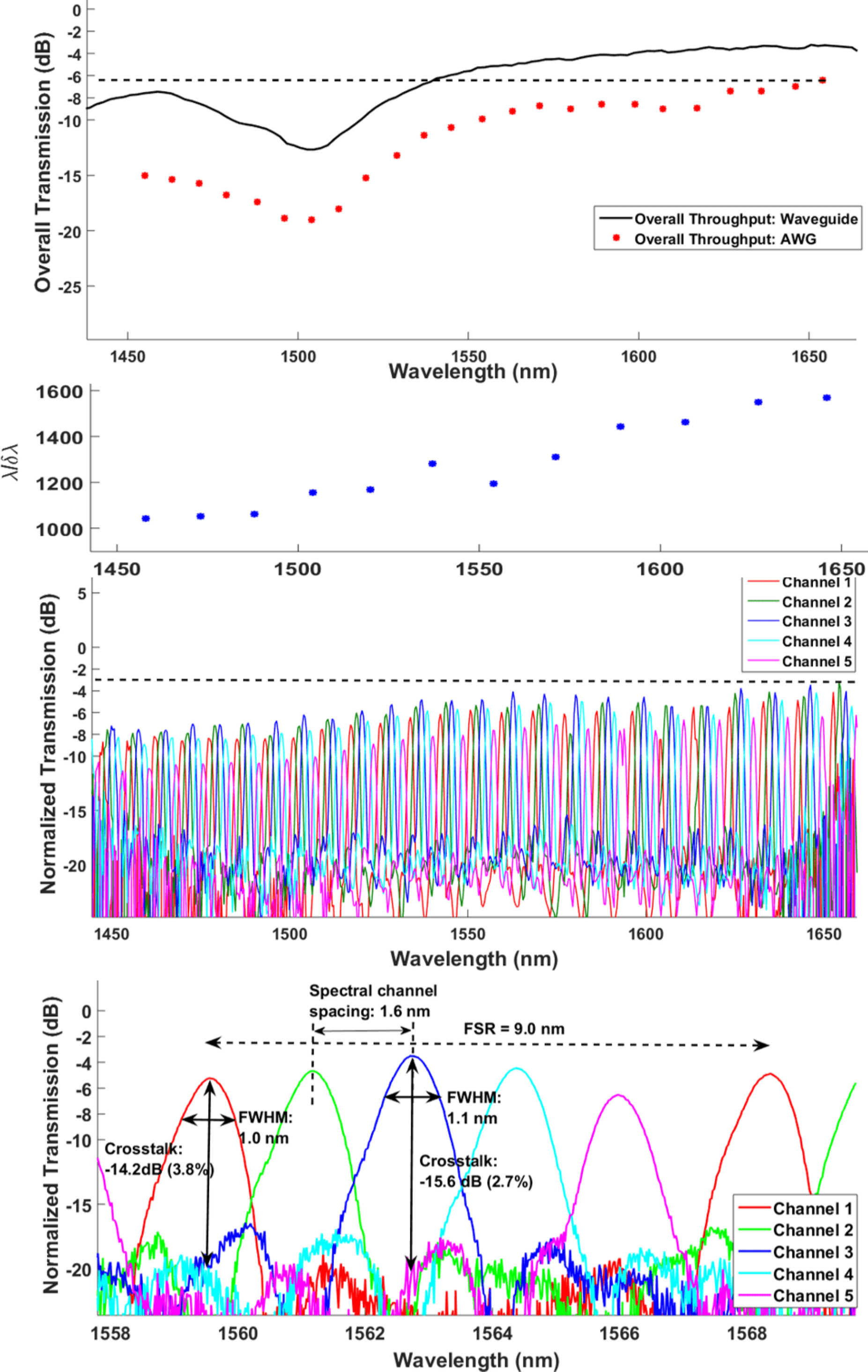}
   \end{tabular}
   \end{center}
   \caption[Taper result]{Panel 1: The overall throughputs of AWG \#1 and the curved reference waveguide are shown. The dashed line indicates the peak overall throughput of the AWG (about $-$6.4 dB or $\sim$23\%). The red dots represent the overall throughputs measured in the centers of the orders. Panel 2: The measured resolving power of the AWG as a function of wavelength. Panel 3: The transmission of the AWG normalized to the curved reference waveguide is shown for all 23 spectral orders. The five colors show the `on-chip throughput' of the five output channels. The dashed line represents the peak on-chip (i.e. normalized to the reference waveguide) throughput of the AWG (about $-$3 dB or $\sim$50\%). Panel 4: A more detailed view of one of the spectral orders is presented to show the FSR, spectral FWHM, spectral channel spacing, and crosstalk. The measurement errors are less than 0.1 dB, so no error bars are shown. \label{fig:Taper Result} } 
   \end{figure}

AWG \#1 was characterized according to the procedure discussed in section 3.3. Figure \ref{fig:Taper Result} shows the overall transmission (fiber-AWG-fiber) of the 2.0 $\mu m$ $\times$ 0.1 $\mu m$ AWG and that of the curved reference waveguide, both with coupling tapers on the input and output ends. The points indicate the transmission for the central wavelengths of each spectral order of the AWG. The second panel shows the on-chip throughput of the AWG (i.e. AWG transmission normalized to the curved reference waveguide). Note that the on-chip throughput is roughly uniformly high over the entire H band. The peak overall transmission is about 23\%(-6.4 dB), which is twice that of our previous AWGs \cite{gatkine2016development}. The fiber-taper coupling efficiency is $\sim$95\%($\sim$0.22 dB) per facet at 1550 nm\cite{zhu2016ultrabroadband} and the extra loss due to curvature of the waveguides is negligible compared to the propagation loss, since the minimum bending radius used is 2.5 mm \cite{bauters2010ultra}. Thus, the propagation loss is about 1.5dB/cm after accounting for coupling and taper losses. The additional AWG loss is roughly 3 dB (50\%) at 1.6 $\mu m$, which is due to additional propagation loss in the curved waveguides of the AWG of $\sim$0.3dB (from the length in excess of reference waveguide) and the loss at the waveguide-FPR interfaces of $\sim$0.7 dB per interface (i.e. 85\% transmission per interface). This is a major contributor to the on-chip loss since there are four such interfaces. 

Therefore, the waveguide-FPR interface taper needs to be further optimized for a better conversion efficiency between the slab mode of the FPR and the waveguide mode. The overall transmission degrades towards shorter wavelengths as a result of scattering from stress-induced microcracks\cite{irene1976residual} and broad absorption features due to hydrogen bonds to silicon, nitrogen, and oxygen in the PECVD SiO$_2$ and LPCVD Si$_3$N$_4$. The broad absorption at 1505 nm due to Si-H bonds is particularly pronounced. We address the absorption problem in the next section.
   
The crosstalk of the AWG is in the range of $-$15 to $-$16 dB at wavelengths longer than 1540 nm. At shorter wavelengths, the crosstalk slowly increases to reach $-$10 dB at 1450 nm. This variation may not be real, since the source power drops by 20 dB at 1450 nm (compared to the power at 1550 nm). Therefore, the crosstalk degradation might simply be due to the Optical Spectrum Analyzer hitting the noise floor. It should be noted that, in an astronomical spectrograph, the spectral FWHM is the important parameter rather than the crosstalk. So a crosstalk of about $-$10 dB is acceptable. The spectral FWHM of the output channels is 1.2 $\pm$ 0.2 nm, which implies a spectral resolving power ($\lambda/\delta\lambda$) of roughly 1300. The measured resolving power as a function of wavelength is shown in the second panel of Fig. \ref{fig:Taper Result}. The non-uniformity among the five channels within a spectral order, as seen in the bottom panel of Fig. \ref{fig:Taper Result}, is due to the intensity envelope of the far-field pattern of the waveguides which illuminates the output FPR. These differences in intensity can be reduced by using suitable mode-field converters at the interface of arrayed waveguides and the output FPR \cite{sakamaki2009loss}. 

\subsection{AWG \#1: annealing}
\begin{figure} [ht]
   \begin{center}
   \begin{tabular}{c} 
   \includegraphics[height=5cm]{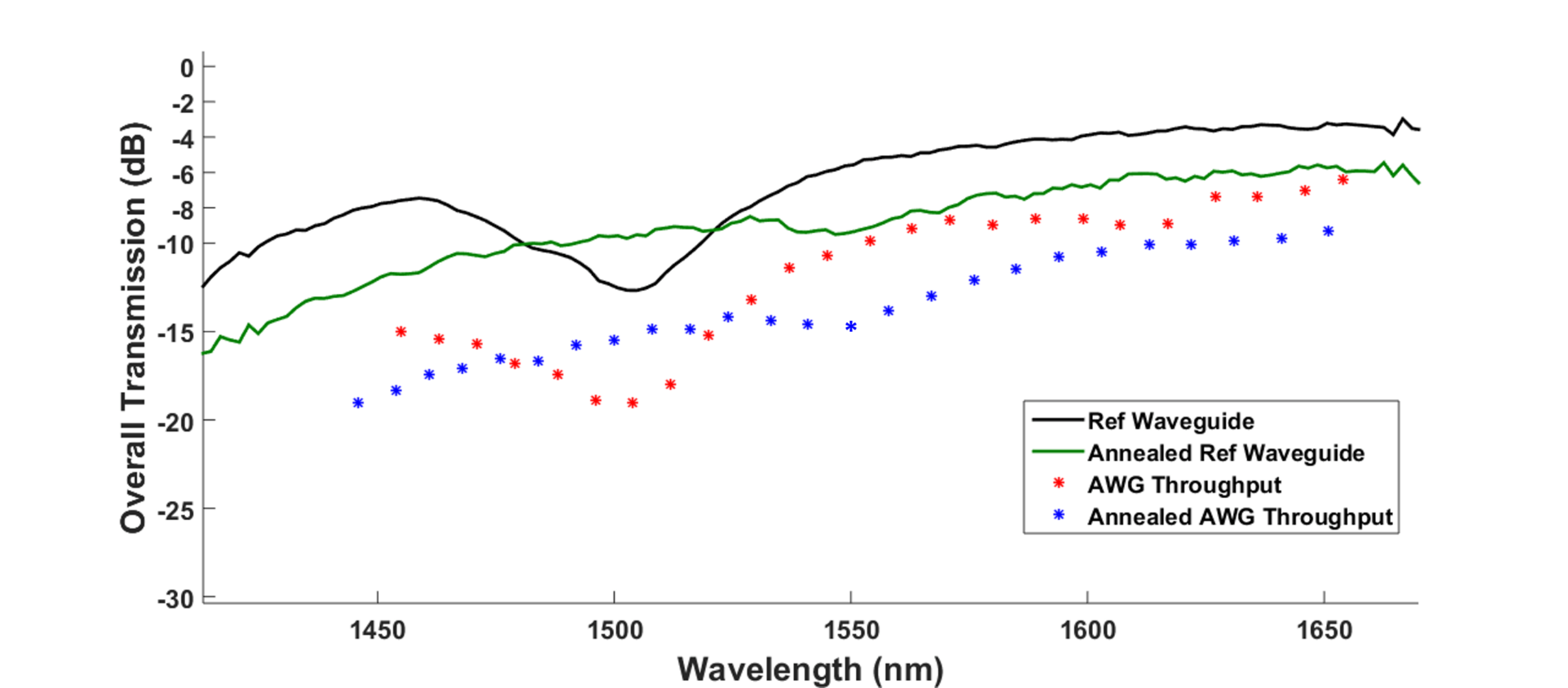}
%
   \end{tabular}
   \end{center}
   \caption[Annealing result] 
   { \label{fig:Annealing result} The overall transmission of AWG \#1 and the curved reference waveguide before and after annealing. The absorption peak near 1500 nm is mitigated to a large extent by the annealing process, but the overall throughput has degraded. The wavelength shift between the order centers of the original and annealed AWGs is due to a change in the effective index of refraction of the waveguides as a result of annealing. 
 }
   \end{figure}

The hydrogen-bonds (especially Si-H) in the PECVD SiO$_{2}$ cladding cause a broad absorption feature seen in the overall transmission of the AWG \cite{henry1987low}. We use high-temperature annealing of the sample to liberate the trapped hydrogen and minimize the absorption. For this, we use a custom recipe of progressively heating the sample in the annealing chamber (in the presence of air) to 800$^\circ$C and 1000$^\circ$C for 30 minutes each, then heating up to 1200$^\circ$C for 2 hours, followed by a progressive cool down. The overall transmission of the annealed sample and comparison with the original AWG (prior to annealing) is shown in Fig.\ref{fig:Annealing result}. The main advantage of the annealed AWG is the improvement of the overall throughput in the 1475$-$1525 nm range. The overall throughput improved from 1.2\% ($-$19.0 dB) at 1504 nm in the original AWG to 3.2\% at 1508 nm in the annealed AWG for the same spectral order. However, at longer wavelengths, a degradation of the overall throughput is observed. The overall throughput dropped from 20\% ($-$7.0 dB) at 1646 nm for the original AWG to 12\% ($-$9.2 dB) at 1651 nm for the annealed AWG for the same spectral order. The propagation loss is estimated to be $\sim$2.6 dB/cm, from the comparison between the reference waveguide transmissions of the original and annealed reference waveguides. Thus, the annealing treatment removes the absorption peak, but it also reduces the overall transmission of the waveguides and the AWG. We believe this is due to the micro-cracks generated by stress along the Si$_3$N$_4$/SiO$_2$ interface at high temperatures. One way to alleviate this issue is to use LPCVD for the deposition of cladding SiO$_{2}$. We plan to explore this problem in the future. As shown in Fig. \ref{fig:Annealing result}, the central wavelengths of the spectral orders also shift towards longer wavelengths by $\sim$4-5 nm due to densification, a thermally induced increase of $\sim$0.3\% in the index of refraction of SiO$_2$ (this increment was calculated by comparing the central wavelength, $\lambda = n_{eff}\Delta L/m$ before and after annealing).      

\subsection{AWG \#2: cleaving at the output FPR}

\begin{figure}[htbp]
\subfloat[]{\includegraphics[height=7cm]{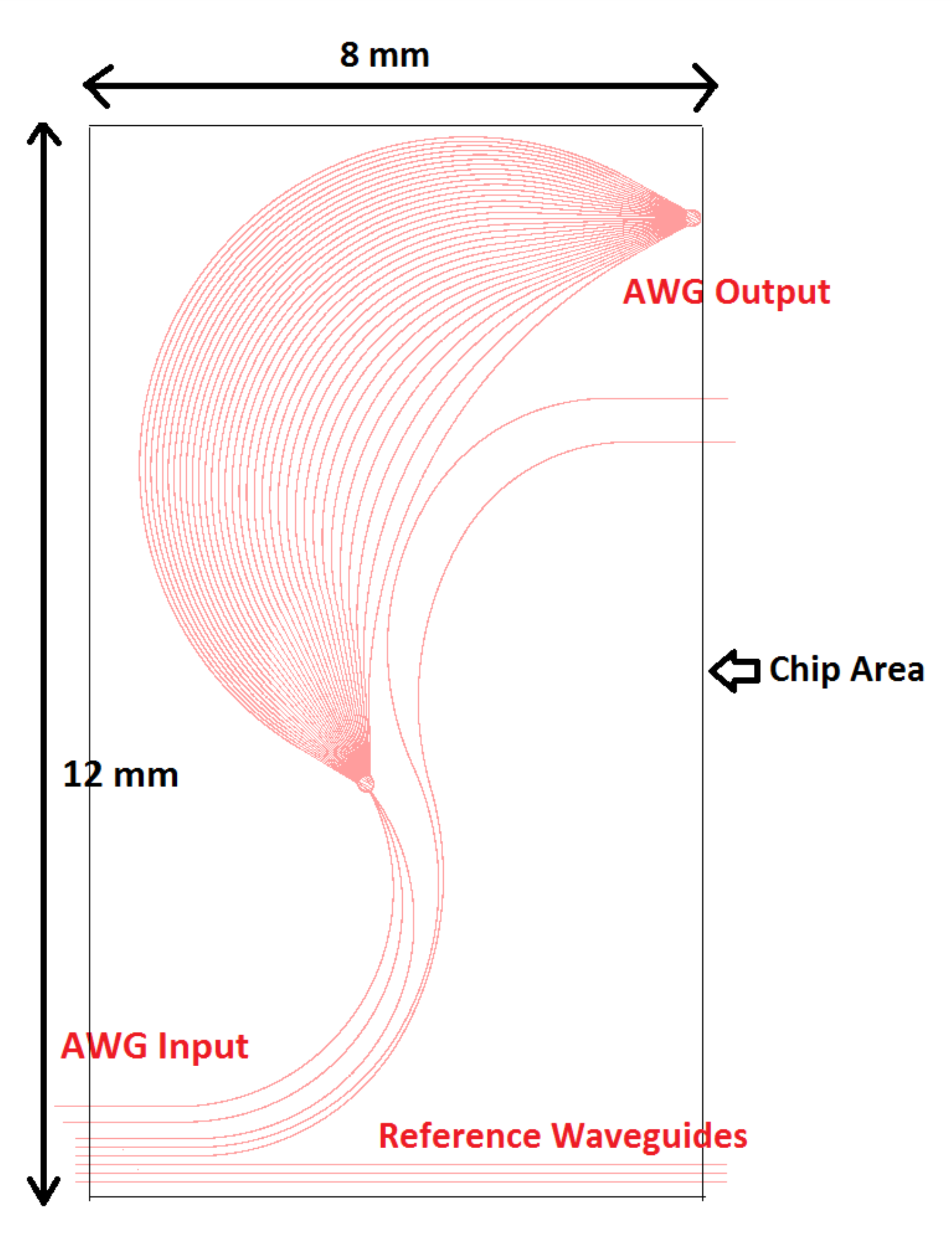}}
\subfloat[]{\includegraphics[height=6cm]{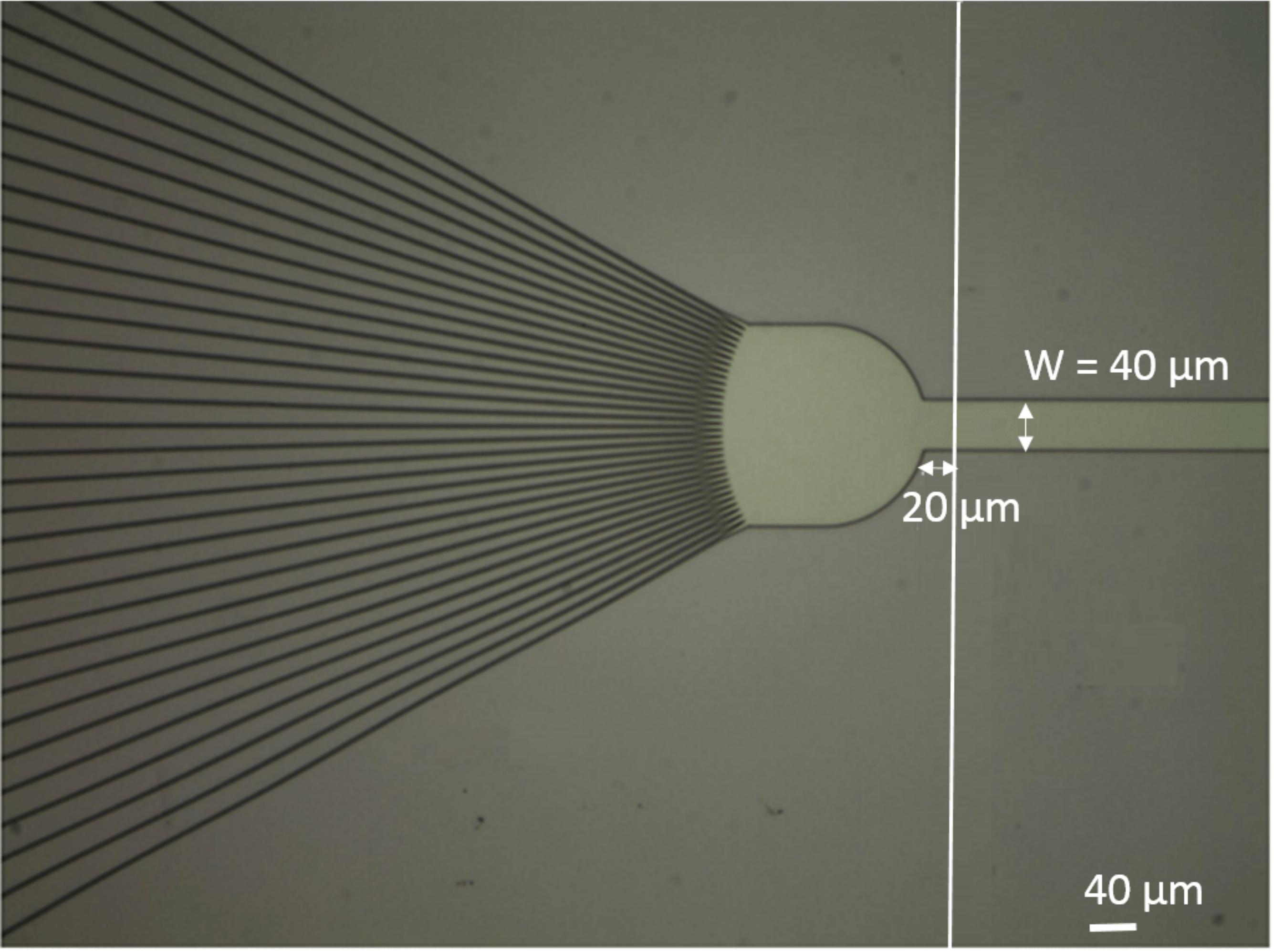} }
\caption{ \label{fig:Modified AWG CAD} a) The modified CAD of AWG \#2 to maintain the cleaving plane parallel to the crystal plane of the wafer. As a result, the device footprint is slightly smaller than Fig. \ref{fig:AWG_CAD}. Out of the three inputs seen in the CAD, only the central input waveguide was used for characterization, the other two are redundant. \label{fig:Cleaving sample} b) A microscope image showing the extra rectangular region added to the output FPR to accommodate the cleaving tolerance of few tens of microns. The sample described in the text and Fig. \ref{fig:cleaved result} was actually cleaved with an unintended 20 $\mu m$ offset (indicated by the vertical white line). }
\end{figure}

For astronomical applications, the AWG needs to be connected to cross-dispersion optics to separate the spectral orders in the perpendicular direction and create a continuous 2D spectrum \cite{cvetojevic2012developing} (unlike the discrete channels of an AWG). For this, the output FPR of the AWG needs to be exposed to illuminate the cross-dispersing optics. In this subsection, we present an AWG chip cleaved at the output FPR and the results of its characterization. The design of this AWG is the same as the first AWG, except that the output waveguides are not present. The fiber-coupling tapers are therefore used only on the input side. The CAD of this AWG is modified so as to have the focal plane of the output FPR along the crystal plane of the wafer for optical-quality cleaving. The length of the input waveguide is kept the same as that in AWG \#1. The modified CAD is shown in Fig. \ref{fig:Modified AWG CAD}. Since the cutting edge has a tolerance of only about 10 $\mu m$, we added an extra rectangle of width 40 $\mu m$ at the end of the FPR to ensure that the light continues to propagate through the nitride region (i.e. FPR slab) even if there is an unintended offset in the cleaving position of a few tens of microns from the focal plane of the output FPR. In case of a cleaving offset, the effect of the extra rectangular section can be assessed by considering the extra length as an effective increment in the length of output FPR slab. This will alter the spectral channel spacing for a spectral order $m$, according to the following relation \cite{takahashi1995transmission}:

\begin{equation}\label{eq:1}
\Delta \lambda = \frac{n_{slab}}{m}\times \frac{D_{i}D_{o}}{L_o},
\end{equation}
where $\Delta \lambda$ is the spectral channel spacing, $L_{o}$ (= 200 $\mu m$ $+$ cleaving offset) is the length of the output FPR, and $D_{i}$ and $D_{o}$ are the waveguide separation at the FPR-output interface (6 $\mu m$) and arrayed waveguide-FPR interface (6 $\mu m$), respectively. Hence, for each spectral order, $\Delta \lambda$ is inversely proportional to $L_{o}$ and $d(L_{o})/L_{o} = - d(\Delta\lambda)/\Delta\lambda$. Also, the focal plane of the AWG (i.e. the output FPR) is on the Rowland circle and therefore curved, but the cleaving happens along a line (i.e. crystal plane). This will cause a phase-mismatch of the interfering waves leading to a distortion of the spectral field pattern.  

The sample was cleaved and had an unintended offset of nearly 20 $\mu m$ (10\% of the length of FPR) from the focal plane of FPR (as shown in Fig. \ref{fig:Cleaving sample}). The AWG was annealed and characterized with a UHNA3 fiber scanning across the output FPR line over a range of 12 $\mu m$ in steps of 2 $\mu m$, with a positional accuracy of 0.1 $\mu m$. The same characterization setup as Fig. \ref{fig:Characterization_stage} was used. 

\begin{figure} [ht!]
   \begin{center}
   \begin{tabular}{c} 
   \includegraphics[height=9.9cm]{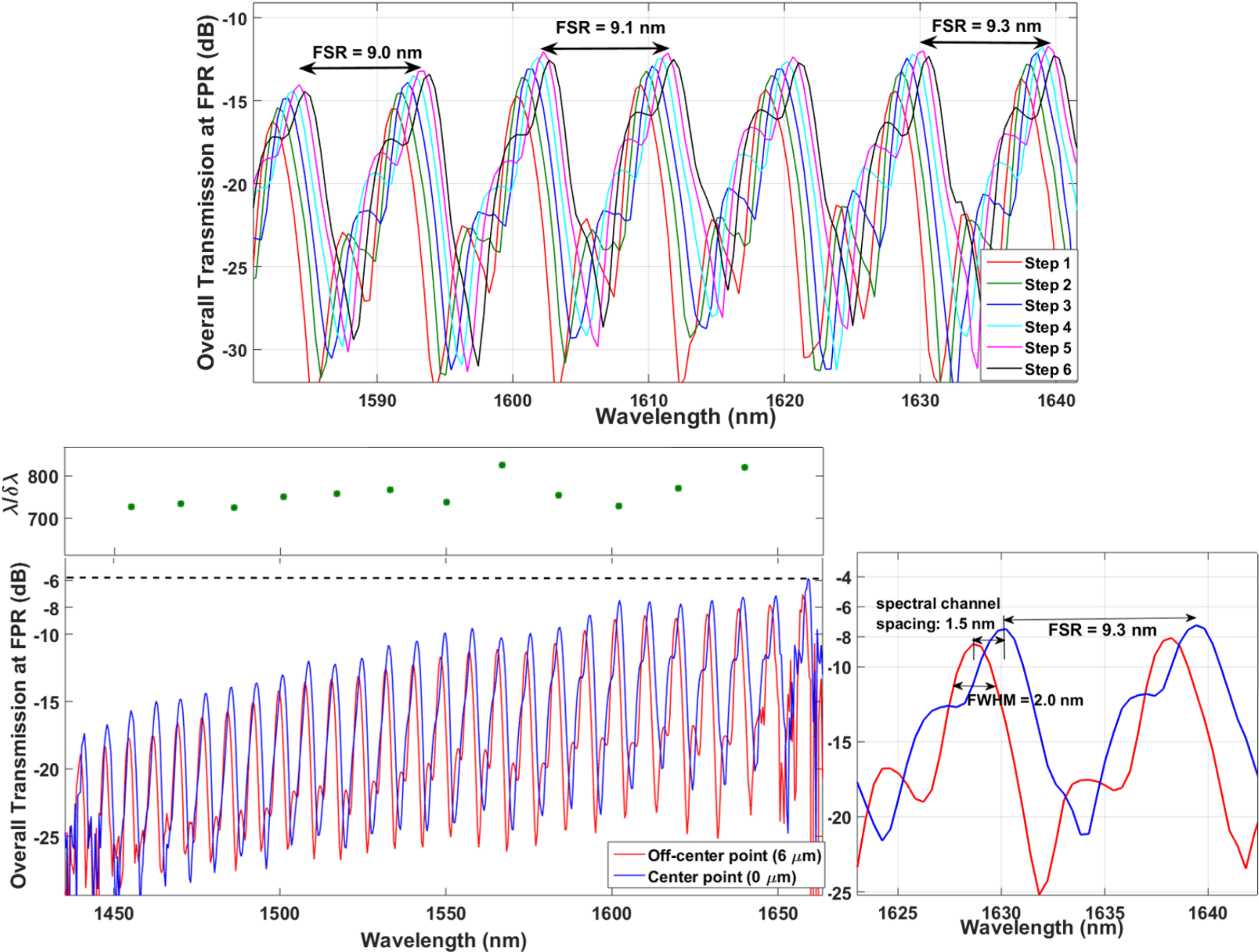}
%
   \end{tabular}
   \end{center}
   \caption[cleaved result] 
   { \label{fig:cleaved result} Top panel: A section of the spectral response (overall throughput) of AWG \#2 at six consecutive points (spaced by 2 $\mu m$) around the center of the FPR. Bottom left panel: The overall throughput integrated over a length of 6 $\mu m$ along the FPR. The blue line shows the integrated throughput for the central point and the red line shows the same for a point 6 $\mu m$ offset from the center. The dashed line indicates the peak overall throughput. The window above shows the measured variation of resolving power as a function of wavelength. Bottom right panel: A zoomed-in view of a section of the 6 $\mu m$ integrated throughput response, showing the FSR, spectral channel spacing, and spectral FWHM.}
   \end{figure}

The top panel of figure \ref{fig:cleaved result} shows a section of the spectral response (overall throughput) at six consecutive points (spaced by 2 $\mu m$) around the center of the FPR. For comparison with the transmission of the annealed AWG \#1, the response needs to be integrated over 6 $\mu m$ (since the spatial output channel spacing of the AWG is 6 $\mu m$, as described in section 3.1). When the fiber samples a region of the FPR, the observed power is the convolution of the mode-profile of the fiber and the spatial distribution of power at the FPR. Therefore, ideally, the fiber response should be de-convolved from the observed power to obtain the spatial distribution of power across the FPR and then it should be integrated over 6 $\mu m$ for accurate comparison. However, such treatment would require a much finer scan with the fiber with spatial steps of the order of 0.5 $\mu m$ across the FPR. This problem can be circumvented if we use a fiber with a narrow mode-size (FWHM) and arithmetically sum the outputs of the sampled sections of the FPR over 6 $\mu m$ to get the integrated power over 6 $\mu m$. The UHNA3 fiber has a narrow mode-FWHM of 1.6 $\mu m$ at 1550 nm. Therefore, it is safe to sum three consecutive steps of 2 $\mu m$ to obtain the integrated power over a 6 $\mu m$ region of the FPR. This integrated power is measured at the center of the scan range (in blue) and also at a point 6 $\mu m$ offset from the center (in red) in the bottom panel of Fig. \ref{fig:cleaved result}.

As mentioned earlier, the added rectangle allows for cleaving offsets, albeit at the cost of introducing distortions of the spectral intensity distribution. The peak throughput at the output face of the FPR is 26\% ($-$5.9 dB). For comparison, the overall throughputs of the annealed AWG \#1 (shown in Fig. \ref{fig:Annealing result}) and annealed-cleaved AWG \#2 are 12\% ($-$9.2 dB) and 19.4\% ($-$7.1 dB), respectively, for the same spectral orders at $\sim$1650 nm. This improvement in throughput can be explained as a result of the absence of the output waveguides (avoiding $\sim$1.5 dB in propagation loss, estimated from section 4.2), the absence of coupling and taper losses ($\sim$0.4 dB), and the replacement of the output FPR-waveguide interface with an FPR-fiber interface. Note that the propagation loss in the annealed AWG was $\sim$2.6 dB/cm as estimated in section 4.2. The throughput for the 6 $\mu m$ offset channel is lower than the central channel by about 1-1.5 dB. A possible reason for the attenuation is the offset between the curved FPR focal plane and the actual cleaving plane, which results in increasingly out-of-focus images at locations far from the center. 

In AWG \#2, the 20 $\mu m$ offset is 10\% of the length of the FPR and therefore results in the spectral channel spacing also being reduced by $\sim$10\% (following eq. \ref{eq:1}). As expected from this, the wavelength separation is 1.5 nm at 1630 nm between the center and the off-center response instead of 1.6 nm for the original AWG (note that 6 $\mu m$ is the spatial separation of the adjacent output channels of the designed AWG, as described in section 3.1). Further, the resolving power of the AWG has degraded as a result of de-focusing of the constructive interference peaks. The spectral FWHM measured in the central channel is 2 nm at a wavelength of 1630 nm, resulting in a resolving power ($\lambda/\delta\lambda$) of 815 instead of $\sim$ 1300 for the original AWG. Also, there are substantial sidelobes in the AWG response, as seen in the bottom right panel of Fig. \ref{fig:cleaved result}. This is possibly due to a phase-mismatch of the interfering waves since the cleaving happens along a straight line instead of the curved focal plane on the Rowland circle. Resolving these problems will require tuning the path differences in the array of waveguides so that the focal plane is flat instead of a Rowland circle \cite{lu2003design}. Despite these issues, the device is made robust to cleaving offsets due to the added rectangular patch, without which the throughput would have substantially dropped due to reflections at the silicon nitride $-$ silion oxide interface in addition to de-focusing distortions. Therefore, the idea of adding the extended rectangle to the end of the FPR is pertinent to planning the next step of cross-dispersion and building an integrated photonic spectrograph.  

\section{Polarization insensitivity}

The results presented in the previous section are for TE polarization. The TM mode is more lossy due to its weakly confined mode profile (with a confinement factor of $\sim$2\%), leading to a substantial curvature loss in the AWG. For faint unpolarized astronomical sources, it is important to have a polarization-insensitive spectrograph to obtain maximum signal-to-noise ratios. The AWGs we presented here are based on rectangular ridge waveguides which make them intrinsically anisotropic and therefore birefringent. The TE and TM mode responses will be offset for any spectral order (by roughly $\Delta n_{eff} / n_{eff} \times \lambda $, where $\Delta n_{eff}$ is the difference between the effective indices of the TE and TM polarizations), unless special measures are taken to make the AWG polarization insensitive. One solution involves inserting a quarter-wave plate at the AWG axis of symmetry to equalize the path difference for the TE and TM modes \cite{takahashi1992polarization}. But this method involves the added complexity of inserting material in the chip, which might incur substantial reflection losses. Another method is to have waveguides with a square cross-section. The fabrication of square waveguides with a high confinement factor (>10\%) will require a thick (> 300 nm) deposition of Si$_3$N$_4$. This might lead to an additional sidewall scattering loss due to the non-uniformities associated with deep etching\cite{bauters2011ultra} and high stress in the deposited layers \cite{irene1976residual,smith1990thick}. Also, a deeper etch would potentially make the thinnest structures in the AWG (eg. taper structures) more fragile, and thus the overall fabrication process more difficult. A better solution would be to achieve polarization independence in waveguides with rectangular or square-like cross-sections that can be fabricated without inducing excessive stress associated with the thicker nitride deposition.  One solution towards a polarization independent design is discussed next.

A polarization independent AWG may be designed by tuning the waveguide geometry to get different spectral orders of TE and TM modes to precisely overlap each other, thereby creating an apparent polarization independent response. Assume a birefringent waveguide with TE mode effective index ($n_{eff,TE}$) and TM mode effective index ($n_{eff,TM}$). Also, in our case, we know that ($n_{eff,TM}$ < $n_{eff,TE}$). Say, at a particular wavelength $\lambda$, the TE mode is in spectral order $m$ and the TM mode is in spectral order $m'$. The AWG has a uniform incremental path difference between adjacent waveguides ($\Delta L$), which has to be the same for both polarizations since it is a fixed spatial length, and is given by \cite{okamoto2010fundamentals}:

\begin{equation}
\Delta L = \frac{m\lambda}{n_{eff,TE}} = \frac{m'\lambda}{n_{eff,TM}} 
\end{equation}

For $m' = m-p$ where $p$ is an integer, we get the condition for order overlap polarization independence, 

\begin{equation}
\frac{n_{eff,TM}}{n_{eff,TE}} = 1-\frac{p}{m}
\end{equation}

\begin{table}[ht]
\caption{Search for appropriate waveguide geometry (integer solution to $p$) for a polarization insensitive AWG for TE order = 165} 
\label{tab:polarization_ind}
\begin{center}       
\begin{tabular}{|c|c|c|c|c|} 
\hline
\rule[-1ex]{0pt}{3.5ex}  Width (nm) & Height (nm) & TE ($n_{eff,TE}$) & TM ($n_{eff,TM}$) & $p$ \\
\hline
\rule[-1ex]{0pt}{3.5ex}  1000 & 100 & 1.4445 & 1.4436 & 0.59   \\
\hline
\rule[-1ex]{0pt}{3.5ex}  1200 & 100 & 1.4494 & 1.4442 & 0.95   \\
\hline
\rule[-1ex]{0pt}{3.5ex}  1400 & 100 & 1.4567 & 1.4452 & 1.30   \\
\hline
\rule[-1ex]{0pt}{3.5ex}  1900 & 100 & 1.4647 & 1.4469 & 2.00  ** \\
\hline
\rule[-1ex]{0pt}{3.5ex}  2600 & 100 & 1.4718 & 1.4492 & 2.54   \\
\hline
\rule[-1ex]{0pt}{3.5ex}  3000 & 100 & 1.4744 & 1.4502 & 2.70   \\
\hline
\end{tabular}
\end{center}
\end{table}

Thus, if we find a waveguide geometry with $n_{eff,TE}$ and $n_{eff,TM}$ such that it gives an integer solution to $p$, then we will essentially get the $m$$^{th}$ order of TE and the $(m-p)$$^{th}$ order of TM overlapping to give an apparent polarization independence. As an example, consider waveguides with a thickness of 100 nm and a width close to 2 $\mu m$ to for a geometry that can give an integer solution to $p$ around our spectral order of 165 at a wavelength of 1600 nm. The search is summarized in Table \ref{tab:polarization_ind}. The solution was found to be $p$ = 2 at width = 1.9 $\mu m$. Therefore, an AWG, constructed with a waveguide geometry of 1.9 $\times$ 0.1 $\mu m$ and designed to have a spectral order ($m$) of 165 at 1600 nm wavelength, will have an apparent polarization independence around 1600 nm due to the overlap of $165^{th}$ TE spectral order and $163^{th}$ TM order. This technique can also be used to attain apparent polarization independence by fixing the waveguide geometry and tuning the spectral order for the overlap. In the future, we will explore this property as a factor in selecting the waveguide geometry for the AWG. One important aspect of this approach is the impact of the refractive index dispersion coefficients for TE and TM modes on the free spectral range (FSR) as a function of wavelength. The FSR is given by:    

\begin{equation}\label{eq: 5}
FSR =  \frac{\lambda}{m}\times\frac{n_{eff}}{n_{group}},
\end{equation}
where $n_{group}$ is the group index ($n_{group} = n_{eff}-
\lambda \frac{dn}{d\lambda}$). Comparing the TE (order $m$) and TM (order $m-p$) FSRs, we get:
\begin{equation}
\frac{(FSR)_{TE}}{(FSR)_{TM}} =  \frac{m-p}{m}\times\frac{n_{eff,TE}}{n_{eff,TM}}\times\frac{n_{group,TM}}{n_{group,TE}} = \frac{m-p}{m}\times\frac{n_{eff,TE}}{n_{eff,TM}}\times\frac{(n_{eff}-
\lambda \frac{dn}{d\lambda})_{TM}}{(n_{eff}-
\lambda \frac{dn}{d\lambda})_{TE}}
\end{equation}

\begin{figure} [ht!]
   \begin{center}
   \begin{tabular}{c} 
   \includegraphics[height=4.5cm]{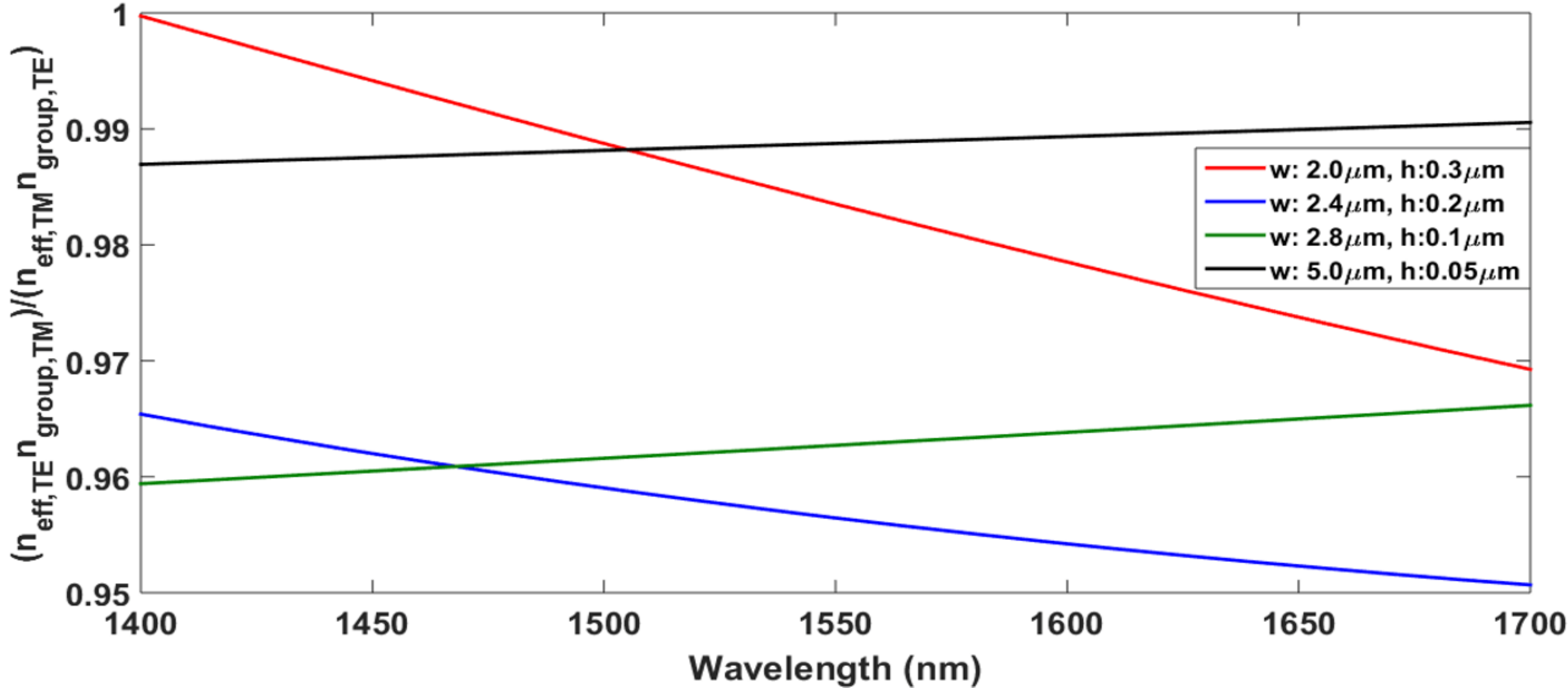}
%
   \end{tabular}
   \end{center}
   \caption[Polarization] 
   { \label{fig:polarization_ratio} The ratio $\frac{n_{eff,TE}}{n_{eff,TM}}\times\frac{(n_{eff}-
\lambda \frac{dn}{d\lambda})_{TM}}{(n_{eff}-
\lambda \frac{dn}{d\lambda})_{TE}}$ (see eqn.\ 5) as a function of wavelength for a set of different single-mode waveguide geometries. A ratio close to unity is desirable to achieve broadband polarization insensitivity over a broad band using the TE-TM order overlap method.}
   \end{figure}

Achieving the polarization insensitivity across several orders would require this ratio to be  close to unity within 1-2\% to avoid any significant offset between TE and TM spectral responses. For materials or geometries yielding small dispersion coefficients, a ratio of FSRs close to unity can be attained for a large m. For large spectral orders (say, $m$ > 100) and small $p$ ($\sim$1), the ratio of FSRs will be determined by $\frac{n_{eff,TE}}{n_{eff,TM}}\times\frac{n_{group,TM}}{n_{group,TE}}$, which needs to be as close to unity as possible. To investigate this ratio, we simulated several single-mode waveguide geometries for a silicon nitride waveguide buried in silica cladding. The simulations were performed using the full-vectorial FDM solver in the FIMMWAVE software \cite{FIMMWAVE}. It was found that for a fixed height, the wider waveguides yielded higher values for this ratio. In Fig. \ref{fig:polarization_ratio}, the calculated ratios for the widest single-mode waveguide geometries are shown for the heights of 50 nm, 100 nm, 200 nm, and 300 nm. It was found that 5000 $\times$ 50 nm and 2000 $\times$ 300 nm are better-suited geometries for this method of polarization independence over a broad band. This technique can also be used for rectangular waveguide geometries and in materials with smaller dispersion coefficients such as SiO$_{2}$.

For non-square waveguides, the mode sizes for TE and TM modes will be different. Consequently, the resolving powers will also be different for the two polarizations, with the larger mode-size polarization leading to a lower resolving power \cite{janz2006frontiers}. Therefore, for a device where polarization insensitivity is attained by using this method, the resulting spectral resolving power will be driven by the polarization with the larger mode size. This simple method to achieve the polarization insensitivity by order overlap can in principle be applied to many other fields of photonic instrumentation with different materials and different geometries. 

\section{Conclusions and future work}

Various techniques of photonics are being widely applied to the field of astronomical instrumentation. This paper describes the fabrication process of new AWGs designed specifically for astronomical applications. These AWGs have peak throughput of $\sim$23\%, resolving power of $\sim$1300, and cover 1450 nm to 1650 nm for TE polarization. The throughput is twice of that obtained in our previous work \cite{gatkine2016development}. These results were obtained using a combination of UHNA3 fiber and an optimized fiber-coupling taper, providing a high coupling efficiency. We further described key practical issues and their possible solutions, such as removing the broad absorption feature around 1500 nm using annealing, cleaving at the focal plane (output FPR) of the AWG to prepare for the cross-dispersion step, and a way to tackle the cleaving tolerances. Finally, a novel way of designing a polarization insensitive AWG without the need for quarter-wave plates was introduced. It is based on the basic idea of tuning the AWG geometry to get different spectral orders of TE and TM modes to precisely overlap each other.   

Future work will focus on testing this idea and improving the overall throughput of our devices. Waveguides with more square-like cross-sections will be fabricated to increase the mode-confinement and reduce the bending loss in TM mode. Our current devices suffer increasingly larger losses at shorter wavelengths. The dominant sources of losses are absorption due to the low quality of PECVD oxide cladding and scattering off of stress-induced micro-cracks. We will explore LPCVD deposition of SiO$_{2}$ to improve the cladding quality and thereby reduce the propagation loss \cite{bauters2011planar}. However, the deposition of a layer of nitride thicker than 1 $\mu m$ with LPCVD often induce micro-cracks \cite{smith1990thick}. Since the mode of our waveguides is confined within $\pm$3 $\mu m$ of the cladding, a combination of LPCVD and PECVD processes might provide the best compromise to improve the quality of the cladding and reduce the propagation loss. To mitigate the scattering loss, we will look into stress-management of the waveguides to alleviate the problem of micro-cracks. Another area that can be improved is the interface between the FPR and the waveguides; novel taper geometries will be investigated to optimize the coupling between the waveguides and the FPR. Once the throughput at shorter wavelengths is optimized, the plan is then to fabricate separate J-band optimized AWGs with the idea of using them in concert with the H-band devices (i.e. splitters will be used to separate the J and H bands portions of the spectrum and feed them to J- and H-band AWGs). There is also an interest in increasing the free spectral range so as to cover each of the H and J bands in $\sim$ 10 spectral orders to subsequently reduce the dispersion power required in the cross-disperser optics for order separation. 

\section*{Funding}
W. M. Keck Foundation grant and Kulkarni Summer Research Fellowship  

\section*{Acknowledgments}

The authors would like to thank the staff at the University of Maryland Nanocenter fabrication laboratory for their help with the various fabrication techniques. We also thank Drs. Tiecheng Zhu and Stuart Vogel for providing useful suggestions.

\end{document}